%% file: main_arXiv.tex
\documentclass[twoside,twocolumn,9pt]{article}
\usepackage[utf8]{inputenc}
\usepackage[T1]{fontenc}
\usepackage{extsizes}
\usepackage[super,sort&compress,comma]{natbib} 
\usepackage[version=3]{mhchem}
\usepackage[left=1.5cm, right=1.5cm, top=1.785cm, bottom=2.0cm]{geometry}
\usepackage{balance}
\usepackage{mathptmx}
\usepackage{sectsty}
\usepackage{graphicx} 
\usepackage{lastpage}
\usepackage[format=plain,justification=justified,singlelinecheck=false,font={stretch=1.125,small,sf},labelfont=bf,labelsep=space]{caption}
\usepackage{float}
\usepackage{fancyhdr}
\usepackage{fnpos}
\usepackage[english]{babel}
\addto{\captionsenglish}{%
  
}
\usepackage{array}
\usepackage{droidsans}
\usepackage{charter}
\usepackage[T1]{fontenc}
\usepackage[usenames,dvipsnames]{xcolor}
\usepackage{setspace}
\usepackage[compact]{titlesec}
\usepackage{hyperref}

\renewcommand{\vec}[1]{\bm{#1}}

\newcommand{\ecoli}{{\it E.~coli}}

\usepackage{amsmath}
\usepackage{amssymb}
\usepackage{bm}
\usepackage{siunitx}
\definecolor{cream}{RGB}{222,217,201}

\DeclareSIUnit{\rpm}{rpm}

\begin{document}

\pagestyle{fancy}
\thispagestyle{plain}
\fancypagestyle{plain}{
\renewcommand{\headrulewidth}{0pt}
}

\makeFNbottom
\makeatletter
\renewcommand\LARGE{\@setfontsize\LARGE{15pt}{17}}
\renewcommand\Large{\@setfontsize\Large{12pt}{14}}
\renewcommand\large{\@setfontsize\large{10pt}{12}}
\renewcommand\footnotesize{\@setfontsize\footnotesize{7pt}{10}}
\makeatother

\renewcommand{\thefootnote}{\fnsymbol{footnote}}
\renewcommand\footnoterule{\vspace*{1pt}%
\color{cream}\hrule width 3.5in height 0.4pt \color{black}\vspace*{5pt}} 
\setcounter{secnumdepth}{5}

\makeatletter 
\renewcommand\@biblabel[1]{#1}            
\renewcommand\@makefntext[1]%
{\noindent\makebox[0pt][r]{\@thefnmark\,}#1}
\makeatother 
\renewcommand{\figurename}{\small{Fig.}~}
\sectionfont{\sffamily\Large}
\subsectionfont{\normalsize}
\subsubsectionfont{\bf}
\setstretch{1.125} 
\setlength{\skip\footins}{0.8cm}
\setlength{\footnotesep}{0.25cm}
\setlength{\jot}{10pt}
\titlespacing*{\section}{0pt}{4pt}{4pt}
\titlespacing*{\subsection}{0pt}{15pt}{1pt}

\fancyfoot{}
\fancyfoot[LO,RE]{\vspace{-7.1pt}\includegraphics[height=9pt]{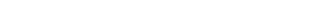}}
\fancyfoot[CO]{\vspace{-7.1pt}\hspace{13.2cm}\includegraphics{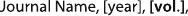}}
\fancyfoot[CE]{\vspace{-7.2pt}\hspace{-14.2cm}\includegraphics{head_foot/RF}}
\fancyfoot[RO]{\footnotesize{\sffamily{1--\pageref{LastPage} ~\textbar  \hspace{2pt}\thepage}}}
\fancyfoot[LE]{\footnotesize{\sffamily{\thepage~\textbar\hspace{3.45cm} 1--\pageref{LastPage}}}}
\fancyhead{}
\renewcommand{\headrulewidth}{0pt} 
\renewcommand{\footrulewidth}{0pt}
\setlength{\arrayrulewidth}{1pt}
\setlength{\columnsep}{6.5mm}
\setlength\bibsep{1pt}

\makeatletter
\newlength{\figrulesep}
\setlength{\figrulesep}{0.5\textfloatsep} 
\newcommand{\topfigrule}{\vspace*{-1pt}%
\noindent{\color{cream}\rule[-\figrulesep]{\columnwidth}{1.5pt}} }

\newcommand{\botfigrule}{\vspace*{-2pt}%
\noindent{\color{cream}\rule[\figrulesep]{\columnwidth}{1.5pt}} }
\newcommand{\dblfigrule}{\vspace*{-1pt}%
\noindent{\color{cream}\rule[-\figrulesep]{\textwidth}{1.5pt}} }

\makeatother
\twocolumn[
\begin{@twocolumnfalse}
{\includegraphics[height=30pt]{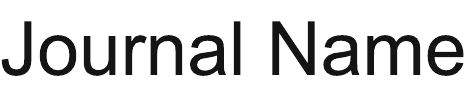}\hfill\raisebox{0pt}[0pt][0pt]{\includegraphics[height=55pt]{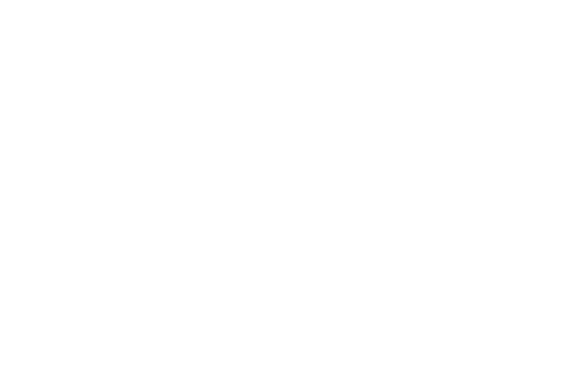}}\\[1ex]
\includegraphics[width=18.5cm]{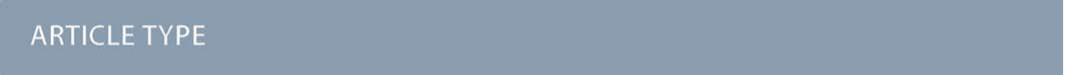}}\par
\vspace{1em}
\sffamily
\begin{tabular}{m{4.5cm} p{13.5cm} }

\includegraphics{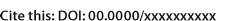} & \noindent\LARGE{\textbf{Elongation suppresses rheotaxis and enables microfluidic enrichment of $\beta$-lactam-resistant bacteria}} \\
\vspace{0.3cm} & \vspace{0.3cm} \\




& \noindent\large{Ran Tao,\textit{$^{a}$} Nathaniel C. Esteves,\textit{$^{b}$} Jay Zhu,\textit{$^{b}$} and Arnold J. T. M. Mathijssen\textit{$^{a}$}}\par\medskip{\large\bfseries\itshape Accepted for publication in \textit{Lab on a Chip}.} \\

\includegraphics{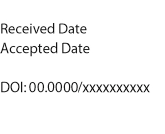} & \noindent\normalsize{Antimicrobial resistance (AMR) complicates the treatment of diseases including lung and urinary tract infections (UTIs), which are among the most common bacterial infections worldwide. 
Motile pathogens can use rheotaxis to swim upstream against fluid flows, potentially promoting access to upper regions of anatomical tracts. However, it remains unclear how antibiotic exposure and resistance influence this transport process. Here, using single-cell tracking microscopy, we investigate how elongation induced by $\beta$-lactam antibiotics affects the rheotactic migration of \ecoli\ in confined microfluidic channels.
Remarkably, we find that rheotaxis can be inhibited 100-fold by antibiotics, even if the susceptible elongated cells remain fully motile. However, resistant bacteria remain short and retain upstream migration under the same conditions. Using genetically engineered bacteria with tunable cell length, we show that the underlying mechanism that governs rheotaxis is the coupling between cell morphology and flow vorticity, where elongated cells are rapidly rotated downstream. Finally, we exploit this length-dependent transport difference to separate short and elongated cells under flow and enrich ampicillin-resistant cells from mixed populations. Together, these results establish bacterial elongation as a key control parameter for rheotactic transport, and provide a proof-of-concept strategy for enriching $\beta$-lactam-resistant bacteria for potential use in rapid AMR detection.}

\end{tabular}

 \end{@twocolumnfalse} \vspace{0.6cm}

  ]

\renewcommand*\rmdefault{bch}\normalfont\upshape
\rmfamily
\section*{}
\vspace{-1cm}


\footnotetext{\textit{$^{a}$Department of Physics \& Astronomy, University of Pennsylvania, Philadelphia, PA 19104, USA; E-mail: amaths@upenn.edu.}}
\footnotetext{\textit{$^{b}$~Perelman School of Medicine, University of Pennsylvania, Philadelphia, PA 19104, USA; junzhu@pennmedicine.upenn.edu.}}



\section*{Introduction}
    Antimicrobial resistance (AMR) is a major global health challenge that threatens the effective treatment of bacterial infections~\cite{WorldHealthOrganization2020TheDeath, UnitedNationsEnvironmentProgramme2023BracingResistance, Ho2025-za}. Among these, urinary tract infections (UTIs) are particularly prevalent and often recur despite antibiotic therapy~\cite{Flores-Mireles2015-fq}. Although substantial progress has been made in identifying the molecular mechanisms underlying AMR, much less is known about how resistance influences the biophysical processes that bacteria use to establish and maintain infections, especially in anatomical tracts subject to fluid flows~\cite{Guasto2012-da, Persat2015-rl, Wheeler2019NotPlankton, Secchi2020-ml, Mathijssen2026BiomedicalFunctionalities}.
    
    A key mechanism that drives infections in such environments is rheotaxis, the directed movement of motile microorganisms against fluid flow~\cite{Hill2007-wr, Kaya2012DirectColi, Figueroa-Morales2015-fx, Mathijssen2019-zy, Figueroa-Morales2020-mb, Jing2020-mj}. In the urinary tract, where urine flow typically serves as a protective barrier, rheotaxis enables bacteria to ascend the tract and reach upstream regions such as the bladder and kidneys~\cite{Flores-Mireles2015-fq,Siryaporn2015-mm}. This behavior has been attributed to a “weathervane effect,” in which the flagellar bundle reorients downstream, aligning cells against the flow~\cite{Hill2007-wr,Mathijssen2019-zy}. Rheotaxis can be enhanced by polymers present in mucus and biofilm matrices~\cite{Torres-Maldonado2024-mp,Cao2024-nm}. 
    \textit{Escherichia coli}, the predominant causative agent of UTIs~\cite{Flores-Mireles2015-fq}, exhibits strong positive rheotaxis, highlighting bacterial motility as an important factor in pathogenesis~\cite{Figueroa-Morales2020-mb, Tao2026-yc, Padron2026-kn}.

    $\beta$-lactam antibiotics, including cephalexin and ampicillin, are commonly used to treat susceptible \ecoli\ infections and inhibit cell-wall synthesis and cell division (Fig.~\ref{fig2}A)~\cite{Kohanski2010-tt,Typas2011-bb}. However, drug concentrations in the urine can drop below the minimum inhibitory concentration (MIC) within hours after treatment~\cite{Melnick2023ImpactInfections}. Under some sub-MIC $\beta$-lactam conditions, susceptible bacteria can continue growing and undergo filamentous elongation rather than being killed, while retaining motility~\cite{Tilanus2023-hf,Maki2000-kp, Cylke2022-iy, Justice2006-gn, Phan2018-gt}.
    
    Despite the clinical importance of AMR and rheotaxis, their interplay remains largely unexplored. Previous studies have shown that antibiotics can alter bacterial metabolism, growth, and swimming behaviors~\cite{Kohanski2010-tt,Cylke2022-iy,Phan2018-gt,Maki2000-kp,Oliveira2022-pc}. However, it remains unclear whether resistance affects the ability of bacteria to migrate upstream against currents and thereby persist in flow environments during antibiotic treatment. Understanding this relationship is essential for fully assessing the risks posed by resistant pathogens.
    
    Here, using a simplified biophysical model inspired by confined flows in the urinary tract, we investigate how $\beta$-lactam-induced elongation affects the upstream migration of \ecoli\ (Fig.~\ref{fig1}). By tracking thousands of individual bacteria, we quantify bacterial upstream migration and resolve the transport dynamics of individual cells under flow. We first show that antibiotic-susceptible cells elongate under $\beta$-lactam exposure while retaining swimming speeds comparable to untreated cells, yet exhibit strongly reduced upstream migration through microstructured environments. In contrast, resistant strains retain a short morphology and preserve upstream migration under antibiotic treatment. To isolate the role of cell length from other antibiotic-induced physiological changes, we use a genetically engineered strain with inducible elongation and demonstrate that elongation alone is sufficient to suppress upstream transport. Finally, we exploit this length-dependent rheotactic response to demonstrate proof-of-concept microfluidic separation and enrichment of ampicillin-resistant cells from mixed populations. Together, these results establish a direct link between bacterial morphology under antibiotic stress and microbial transport dynamics in flow~\cite{Karita2022-tn,Shuppara2025-xl, Chopra2022-ee, Daddi-Moussa-Ider2020-yk}.

\section*{Results}

    \subsection*{Conceptual and experimental framework}
    
    Figure~\ref{fig1} summarizes our investigation of how antibiotic-induced changes in cell length affect upstream transport and illustrates potential future applications. In the urinary tract, bacterial rheotaxis may support upstream migration against urine flow, while oral treatment with $\beta$-lactam antibiotics exposes susceptible cells to conditions that can inhibit division and induce elongation~\cite{Flores-Mireles2015-fq,Siryaporn2015-mm,Kohanski2010-tt,Typas2011-bb,Melnick2023ImpactInfections} (Fig.~\ref{fig1}A, B). This clinical context leads us to investigate how antibiotic-induced elongation alters bacterial transport under controlled flow conditions. We first establish the antibiotic-dependent differences in cell length among the populations used in this study: untreated wild-type cells remain short, antibiotic-susceptible cells elongate under $\beta$-lactam exposure, and antibiotic-resistant cells remain short under treatment (Figs.~\ref{fig1}C and \ref{fig2}C). We next compare short and elongated cells under identical imposed-flow conditions. Short cells sustain near-surface upstream motion, whereas elongated cells undergo more frequent flow-induced reorientation, surface detachment, and downstream advection (Fig.~\ref{fig1}D). To determine whether cell length alone is sufficient to account for this loss of upstream transport, we use a genetically engineered strain with inducible elongation~\cite{Gao2026-vq}. This system allows us to isolate the effect of cell length from other antibiotic-induced changes and quantify how elongation alters the breakout, propagation, and infiltration stages of upstream invasion (Fig.~\ref{fig3}).

    We next test whether this length-dependent transport difference can be used for cell separation. In mixed populations of short and elongated cells, short cells progressively accumulate upstream, whereas elongated cells remain predominantly downstream. We then apply the same principle to mixed populations of ampicillin-resistant and ampicillin-susceptible cells, enriching the resistant population through rheotactic transport (Fig.~\ref{fig1}E). This approach complements broader efforts to use microfluidic transport and physical cell properties for bacterial separation~\cite{Liu2022-uv,Masaeli2012-pq,Tang2022-da,Cruz2019-yx,Zhang2023-ww}.
    
    Thus, the study progresses from antibiotic-dependent differences in cell length to a stage-resolved transport mechanism and finally to flow-mediated enrichment. Figure~\ref{fig1}F illustrates a possible future extension in which enriched cells could be counted without microscopy using impedance cytometry and subsequently analyzed by sequencing or other molecular methods~\cite{Tang2022-kx,Zhu2024-de,Anjum2017-hu}. These downstream readout and identification steps remain conceptual and were not integrated into the present study.

    \subsection*{Antibiotics induce elongation without impairing motility}

    To quantify antibiotic-induced changes in cell length and motility, we exposed wild-type K-12 \ecoli\ to the $\beta$-lactam antibiotics cephalexin and ampicillin. We measured cell length after growth in nutrient-rich medium with and without antibiotics and quantified swimming speed after resuspension in Berg's motility buffer. Representative schematics of cell morphology under each condition are shown in Fig.~\ref{fig2}B, with cell lengths quantified in Fig.~\ref{fig2}C.
    
    Wild-type \ecoli\ cells have an average cell body length of \SIrange{2}{3}{\micro\metre} during their exponential growth phase. Treatment of susceptible cells with cephalexin- or ampicillin-supplemented TB medium for \SI{2}{\hour} inhibited cell division and increased the average cell length to approximately \SI{9}{\micro\metre} (Fig.~\ref{fig2}C; Fig.~S1D, G), consistent with previous observations~\cite{Maki2000-kp,Cylke2022-iy}.
    To isolate the effect of cell length from other antibiotic-induced physiological changes, we used a genetically engineered strain with tunable cell length~\cite{Gao2026-vq}.
    In this strain, elongation is directly controlled through arabinose-inducible expression of SulA from the P$_{\mathrm{BAD}}$ promoter. SulA sequesters FtsZ and inhibits septum formation, thereby blocking cell division~\cite{Higashitani1995-qn}. This engineered strain reached lengths similar to those of antibiotic-treated cells (Fig.~\ref{fig2}C). 
    In contrast, resistant strains retained their ability to divide and remained short under antibiotic treatment.
    An independent phenotypic growth assay further confirmed that the susceptible strain showed no detectable growth at ampicillin concentrations of \SI{16}{\micro\gram\per\milli\liter} or higher, whereas the resistant strain maintained growth through \SI{512}{\micro\gram\per\milli\liter} (Fig.~S1J, K).
    
    Subsequently, bacterial swimming speeds for each condition were measured by tracking individual cells after resuspension in Berg's motility buffer and calculating the trajectory-averaged swimming speed for trajectories lasting longer than \SI{1}{\second} (Fig.~\ref{fig2}D; Fig.~S1E, H). Despite the differences in cell length, swimming speeds remained comparable across conditions, indicating that elongation did not substantially reduce swimming speed under the tested conditions, consistent with previous studies~\cite{Kamdar2023-eb,Maki2000-kp}.
    
    \subsection*{Antibiotics suppress upstream migration in susceptible but not resistant cells}
    
    To study upstream migration in a controlled confined-flow environment inspired by the urinary tract, we fabricated a microfluidic device consisting of two reservoirs connected by a straight channel of width $W=\SI{50}{\micro\metre}$ and depth $H=\SI{10}{\micro\metre}$ (Fig.~\ref{fig2}E). The downstream reservoir represents regions contaminated with bacteria, such as the opening of the urinary tract or a catheter bag, whereas the upstream reservoir represents sterile regions, such as the bladder or kidneys. 
    
    First, under the antibiotic-free control condition, nutrient-rich TB medium was flushed through the device while wild-type \ecoli\ cells were introduced into the downstream reservoir. Under imposed flow, the bacteria readily migrated upstream and accumulated in the upstream reservoir (Fig.~\ref{fig2}F, green), consistent with previous reports~\cite{Figueroa-Morales2015-fx,Tao2026-yc}. The first invading cell typically reached the upstream reservoir within \SI{60}{\second} after traversing the \SI{2000}{\micro\metre} channel against the imposed flow (Fig.~S2A, B).
    During the course of the experiments, each lasting up to \SI{8}{\hour}, the cells continued to grow and divide in the presence of nutrients, leading to a persistent flux of bacteria moving upstream. 
    Consequently, cells progressively accumulated in the upstream reservoir (Fig.~S2C).
    All bacteria within the channel and the upstream reservoir consistently exhibited a short morphology.
    
    To mimic infection treatment with antibiotic exposure, we flushed the microfluidic device with TB medium supplemented with cephalexin (Fig.~\ref{fig2}F, magenta) or ampicillin (red). Under both antibiotic conditions, upstream migration by susceptible cells was strongly suppressed. During the first \SI{30}{\minute} of antibiotic exposure, some cells remained short and were able to reach the upstream reservoir. As the antibiotic effect became established, the cells elongated and the upstream migration progressively decreased until it was completely abolished (Fig.~S2D--F). Most susceptible cells within the device showed elongated morphologies.
    
    High-magnification inspection of the upstream reservoir after \SI{8}{\hour} revealed a dense bacterial accumulation under the antibiotic-free condition (green), while the accumulation was approximately 100-fold reduced when wild-type (susceptible) bacteria were exposed to cephalexin (magenta) or ampicillin (red) (Fig.~\ref{fig2}G, H). In contrast, ampicillin-resistant cells (yellow) retained a short morphology under ampicillin treatment and showed unhindered upstream migration.
    
    \subsection*{Cell elongation impairs all stages of upstream invasion}

    In separate experiments, short wild-type cells and genetically engineered cells with induced elongation were inoculated into the downstream reservoir, while a steady flow of motility buffer was applied (Fig.~\ref{fig3}A). Under these conditions, short cells consistently migrated upstream and accumulated in the upstream reservoir, while elongated cells did not reach the upstream reservoir and remained trapped near the downstream entrance (Fig.~\ref{fig3}B).
    
    To resolve this upstream invasion process in more detail, we decomposed it into three stages: (\textit{I}) breakout from the downstream reservoir, (\textit{II}) propagation upstream through narrow channels, and (\textit{III}) infiltration into the upstream reservoir (Fig.~\ref{fig3}A). We introduced this three-stage framework and device geometry in our previous work~\cite{Tao2026-yc}. Here, we use the framework to determine how cell elongation modifies each transport stage. We therefore quantified stage-specific transport metrics as functions of flow strength ($V_{max}$) and cell length ($L$): breakout probability, mean longitudinal propagation velocity, and infiltration probability.
    
    During the breakout stage (\textit{I}) (Fig.~\ref{fig3}C, D; Fig.~S3), wild-type cells were able to enter the channel mouth. As expected, breakout events became less frequent under stronger flows, although occasional successes occurred when multiple cells accumulated near the entrance. In contrast, elongated cells exhibited near-zero breakout probability across the tested flow conditions. Their larger aspect ratio increases their susceptibility to surface detachment in regions with high streamline curvature near the channel entrances, resulting in near-zero breakout probabilities for elongated bacteria.
    
    During the propagation stage (\textit{II}) (Fig.~\ref{fig3}E, F; Fig.~S4), wild-type cells swam upstream along the channel edges but were advected downstream when they detached from the surfaces and moved into the faster flows near the channel centerline. At weak shear rates, the upstream swimming distances exceeded the downstream drift, resulting in net upstream migration. At stronger shear, detachment events dominated and wild-type cells were carried downstream. In contrast, elongated cells spent little time near the edges, detached rapidly, and were frequently advected downstream. Even under weak flow conditions, they exhibited net downstream velocities, and at stronger flows they accumulated near the channel centerline, where they were transported downstream faster than wild-type cells. We compared these experiments with a three-dimensional rheotaxis model that describes bacterial motion in thin rectangular microfluidic channels (Fig.~\ref{fig3}E; Notes S8 and S9). In the model, short and elongated cells were simulated under identical conditions, except that the cell-body aspect ratio was increased from the wild-type-like range \(\Gamma=2\text{--}4\) to the elongated range \(\Gamma=8\text{--}10\). The simulations reproduced the qualitative features of the experimental trajectories: short cells maintained stable edge-following motion and achieved net upstream propagation, whereas elongated cells detached more frequently from the channel edges and were advected downstream. By computing the ensemble-averaged longitudinal velocity \(\langle V_x\rangle\) as a function of flow strength, we found that the model (Fig.~\ref{fig3}, lines) captures the experimentally observed differences (Fig.~\ref{fig3}, points), with short cells exhibiting net upstream migration and elongated cells undergoing downstream transport. The agreement between experiments and simulations supports the conclusion that cell elongation alone, through enhanced sensitivity to local shear and vorticity, is sufficient to impair upstream propagation during migration.
    
    During the final infiltration stage (\textit{III}) (Fig.~\ref{fig3}G, H; Fig.~S5), wild-type cells crossed the curved streamlines at the channel exit and successfully entered the upstream reservoir. In contrast, elongated cells exhibited near-zero infiltration probability across the tested flow conditions, consistent with their inability to escape from downstream openings in the first stage.
    
    Together (\textit{I}, \textit{II}, \textit{III}), these results demonstrate that cell elongation suppresses upstream migration across all three stages of the invasion process. Elongated cells show increased sensitivity to local vorticity ($\vec{\omega} = \vec{\nabla}\times\vec{v}$), particularly near channel entrances, exits, and bounding surfaces. Frequent detachment from the channel edges prevents elongated cells from sustaining upstream migration, thereby reducing their ability to reach upstream regions. In contrast, short cells maintain stable edge-following trajectories and effective rheotaxis.
    
    \subsection*{Rheotaxis-driven separation of bacterial populations by cell length}
    
    This finding, that bacterial cell length strongly influences their upstream swimming performance, can be exploited for cell separation under flow. 
    To test this prediction, we mixed short cells (green) and elongated cells (blue) and introduced the population into the downstream reservoir of the microfluidic device (Fig.~\ref{fig4}A). 
    Over time, the two populations spatially segregated. 
    Short cells efficiently migrated through the channel and accumulated in the upstream reservoir. 
    In contrast, elongated cells did not propagate beyond the channel entrance and remained trapped near the downstream opening, where they gradually accumulated.
    
    This separation is robust and reproducible, arising from fundamental differences in how cell length interacts with shear flow and channel geometry. Short cells maintain stable edge-following trajectories and resist detachment from channel walls, whereas elongated cells detach more readily and are advected downstream, preventing sustained upstream migration. The imposed flow therefore acts as a hydrodynamic selector that partitions cells according to their length-dependent rheotactic behavior~\cite{Liu2022-uv,Masaeli2012-pq,Tang2022-da,Cruz2019-yx,Zhang2023-ww}.

    To confirm that this separation arises from flow-driven effects rather than geometric features of channel openings, we performed control experiments in which both short and elongated cells were introduced into the microfluidic device in the absence of flow (Fig.~S6A). Under these conditions, cells remained uniformly distributed without preferential accumulation, demonstrating that the observed separation is driven by rheotaxis rather than intrinsic differences in motility or passive transport (Fig.~S6B, C).

    To further demonstrate that the enriched short-cell population can be physically collected, we modified the microfluidic design by adding a lateral outlet channel connected to the upstream reservoir (Fig.~S7A). In this configuration, cells reaching the upstream reservoir are redirected into the lateral outlet channel by the imposed flow for collection. When mixed short and elongated cells were introduced into the downstream reservoir, short cells migrated upstream through the device, accumulated in the upstream reservoir in minutes, and were subsequently redirected into the lateral outlet channel (Fig.~S7B). In contrast, elongated cells remained predominantly near the downstream reservoir. This collection process persisted throughout the duration of the experiment (\SI{1}{\hour}), showing that upstream-migration-based enrichment can be directly coupled to physical cell collection. Thus, the device enables the physical collection of the upstream-enriched short-cell population, which could support future downstream culture or molecular analysis.
    
    In the context of antimicrobial resistance, this separation of short and long cells highlights an important distinction: resistant strains retain short morphology under antibiotic treatment and therefore continue to migrate upstream and reach upstream regions. In contrast, susceptible strains elongate and are physically excluded from upstream regions. This hydrodynamic segregation not only underscores the risks associated with antimicrobial resistance, but also suggests a strategy to enrich resistant populations from mixed samples.
    
    \subsection*{Rheotaxis-based enrichment of ampicillin-resistant cells}

    The separation of short and elongated cells suggested that antibiotic-dependent differences in cell length could be exploited to enrich resistant cells from mixed populations (Fig.~\ref{fig5}A). Figure~\ref{fig5}A combines the experimentally demonstrated ampicillin proof of concept with several possible future extensions of the platform. In the present study, we tested only mixed ampicillin-resistant and ampicillin-susceptible \ecoli\ populations after exposure to ampicillin. Patient-derived samples, parallel testing under multiple antibiotic conditions, impedance-cytometry readout, and downstream sequencing are shown as conceptual extensions and were not integrated into the current study.
    
    To demonstrate this enrichment principle experimentally, we cultured an ampicillin-resistant strain together with a wild-type susceptible strain in an ampicillin-supplemented TB medium. After approximately \SI{2}{\hour} of antibiotic exposure, the mixture was suspended in motility buffer before introducing it into the downstream reservoir of the microfluidic device (Fig.~\ref{fig5}B). During the subsequent \SI{30}{\minute} microfluidic selection step, resistant cells (yellow) migrated into the upstream reservoir, whereas susceptible elongated cells remained trapped downstream (Fig.~S8A). Consistent with this separation, \SI{95}{\percent} of the cells accumulating in the upstream reservoir were shorter than \SI{4.6}{\micro\meter} (Fig.~S8B), indicating strong enrichment of the short-cell population.

    To evaluate enrichment performance, we prepared mixed populations with initial resistant-cell fractions of $1\%$, $10\%$, and $50\%$ and quantified the cells accumulating in the upstream reservoir during a \SI{30}{\minute} selection period (Fig.~\ref{fig5}C; Supplementary Note S6). In all three conditions, the total number of cells in the upstream reservoir increased over time. The number of resistant cells accumulating upstream increased with the initial resistant-cell abundance, whereas relatively few susceptible elongated cells reached the upstream reservoir. Consequently, the fraction of resistant cells increased in the upstream population, even when resistant cells were initially rare. At \SI{30}{\minute}, we quantified resistant and susceptible cell counts in the upstream reservoir and calculated the corresponding resistant-cell fraction for each input composition. These measurements demonstrate reproducible enrichment across all tested input compositions. For each input composition, $n=3$ independent experiments were performed using separate samples prepared on different days.

    Together, these results establish a proof-of-concept microfluidic enrichment strategy. The approximately \SI{30}{\minute} timescale refers specifically to the microfluidic selection step following antibiotic exposure. The present experiments quantify enrichment by fluorescence microscopy. Future implementations could integrate impedance cytometry for microscopy-free cell counting and couple the enriched population to sequencing or other molecular analyses.

\section*{Conclusions}
    Our study establishes a direct link between antibiotic-induced cell elongation and the ability of bacteria to migrate upstream against flows. By combining microbiology experiments, microfluidic environments, and computational analysis, we show that $\beta$-lactam-exposed susceptible \ecoli\ elongate and become more prone to detachment in structured flow environments. This elongation suppresses breakout, propagation, and infiltration during upstream migration in a simplified microfluidic system motivated by bacterial transport in the urinary tract. In contrast, resistant cells retain a short morphology and preserve upstream migration under ampicillin treatment. These findings show that resistance not only enables survival under antibiotic exposure but also preserves transport behaviors that facilitate movement against flow.
    
    This perspective complements conventional views of antimicrobial resistance that primarily focus on biochemical and genetic mechanisms. Previous studies have shown how antibiotics alter metabolism, gene regulation, and cell division, and, in some cases, bacterial motility~\cite{Kohanski2010-tt,Anjum2017-hu,Cylke2022-iy,Phan2018-gt}. Our results extend these insights by incorporating hydrodynamic effects, demonstrating how antibiotic-induced elongation influences transport in confined flows. Notably, our results show that elongation preserves intrinsic motility but compromises upstream transport in structured flow environments by increasing susceptibility to detachment and downstream advection. Therefore, transport against flow critically depends on the interplay between cell morphology, motility, and hydrodynamic forces under antibiotic stress~\cite{Mathijssen2019-zy,Tao2026-yc,Tokarova2021-wn}.
    
    The biological implications of this work extend beyond urinary tract infections. Bacterial upstream migration may also be relevant in other pathologies, including lung infections, gastrointestinal diseases, cardiovascular conditions, and the invasion of bacteria into biomedical devices~\cite{Siryaporn2015-mm,Tao2026-yc, Persat2015-rl,Secchi2020-ml,Shen2012-ji}.
    Moreover, bacteria often encounter antibiotics in natural flow environments, such as soil, where antimicrobial agents produced by plants and fungi are transported by rainwater.
    Although more studies are needed to fully investigate the ecological and clinical relevance of our work, our findings provide a universal physical mechanism that could contribute to bacterial persistence and colonization under flow.
    
    Whereas this study focuses on \ecoli, the interplay between cell morphology and rheotaxis likely extends across a wide range of microorganisms. Rheotaxis has been observed not only in swimming bacteria, but also in species that rely on alternative motility modes, including twitching motility driven by type IV pili in \textit{Pseudomonas aeruginosa} and \textit{Xylella fastidiosa}, as well as gliding motility in \textit{Mycoplasma}~\cite{Shen2012-ji,Meng2005-cf,Nakane2022-sm,Rosengarten1988-wh}. While these modes of locomotion typically occur at lower speeds than flagellar swimming, they similarly depend on interactions between cell shape, surface contact, and flow. Our results therefore suggest a general principle: morphological changes, irrespective of the mechanism underlying these changes, can modulate rheotactic transport by altering susceptibility to detachment and advection.
    
    The length-dependent transport difference also provides a physical basis for microfluidic cell enrichment. Under imposed flow, bacterial populations segregate according to cell length, effectively creating a hydrodynamic filtering mechanism. We leveraged this principle to enrich ampicillin-resistant cells from mixed populations using a \SI{30}{\minute} microfluidic selection step following approximately \SI{2}{\hour} of antibiotic exposure. These results demonstrate rapid physical enrichment following antibiotic exposure and motivate integration with downstream phenotypic or molecular readouts.

    The platform also offers practical advantages over many existing microfluidic approaches for bacterial sorting and phenotypic antimicrobial susceptibility testing~\cite{Perez-Rodriguez2022-bg,Zhou2019-gh,Gurung2020-ah,Liu2022-uv,Liu2016-jp,Postek2022-fe}. The current device operates in single-phase flow without requiring multiphase handling. Because separation arises from hydrodynamic interactions rather than geometric exclusion alone, future devices could incorporate parallel channels to increase throughput. These features support the future development of the platform as an enrichment module that could be coupled to independent phenotypic or molecular readouts for rapid AMR detection.

    Further studies could evaluate the platform under more physiologically relevant conditions, including urine or urine-mimicking media, clinically relevant \ecoli\ isolates, and additional urinary tract flow environments. It would also be valuable to test a broader range of $\beta$-lactam antibiotics and resistance mechanisms, as well as to determine how strain-dependent differences in motility, adhesion, and elongation influence enrichment performance. Finally, integrating microscopy-free readouts such as impedance cytometry and coupling the device to downstream culture-based or molecular analysis could further develop the platform toward practical diagnostic use.
    
    In summary, we show that antibiotic-induced elongation suppresses rheotactic transport and that this physical difference can be used to enrich ampicillin-resistant cells from mixed populations. More broadly, these results highlight the importance of linking microbial physiology with hydrodynamics to better understand bacterial behavior in complex environments~\cite{Persat2015-rl,Guasto2012-da,Aranson2022-lz}.

\section*{Methods}
    \subsection*{Bacterial strains and culturing conditions}
    
    The bacterial strain used in this work was \ecoli\ K-12 strain EPB47, an MG1655 derivative carrying a chromosomal \textit{ompA-cfp} fusion (a gift from Mark Goulian, University of Pennsylvania). Cells from frozen stocks were grown at \SI{32}{\degreeCelsius} on lysogeny broth (LB) agar plates (\SI{1}{\percent} Bacto tryptone, \SI{0.5}{\percent} yeast extract, \SI{1.0}{\percent} NaCl and \SI{1.5}{\percent} agar) or in liquid LB with shaking at \SI{250}{\rpm}.

    \subsection*{Induction of cell elongation via controlled \textit{sulA} expression}

    Tunable cell length in \ecoli\ was achieved via heterologous expression of \textit{sulA}, which encodes a repressor of Z-ring formation and inhibits cell division~\cite{Higashitani1995-qn,Gao2026-vq}. The \textit{sulA} gene was PCR-amplified from wild-type \ecoli\ and cloned into the EcoRI and HindIII sites of pBAD24 following standard restriction-enzyme cloning procedures. Use of EcoRI ensured that \textit{sulA} was positioned proximal to the strong ribosome-binding site in the pBAD24 multiple cloning site. The resulting construct was confirmed by Sanger sequencing. \ecoli\ was transformed with the plasmid via chemical transformation and plated on LB agar supplemented with \SI{100}{\micro\gram\per\milli\liter} ampicillin. All growth media were supplemented with \SI{0.2}{\percent} D-glucose to repress \textit{sulA} expression. To induce elongation, glucose was omitted and replaced with \SI{0.2}{\percent} L-arabinose to activate expression from the P\textsubscript{BAD} promoter.
    
    Cultures were grown overnight at \SI{32}{\celsius} with shaking at \SI{250}{\rpm}. A \SI{100}{\micro\liter} aliquot was diluted $10^{-2}$ into fresh TB medium and grown for approximately \SI{5}{\hour} until OD$_{600}$ reached 0.1. For elongation induction, cultures were supplemented with \SI{0.2}{\percent} L-arabinose and incubated for \SI{1}{\hour}, with induction duration determining the average cell length. Cells were then diluted $10^{-2}$ into motility buffer (MB: \SI{0.1}{mM} EDTA, \SI{0.001}{mM} L-methionine, \SI{10}{mM} sodium lactate, \SI{67}{mM} NaCl, \SI{6.2}{mM} K$_2$HPO$_4$, \SI{3.9}{mM} KH$_2$PO$_4$) supplemented with \SI{0.08}{\gram\per\milli\liter} L-serine and \SI{0.03}{\percent} polyvinylpyrrolidone (PVP). The motility buffer was adjusted to pH 7.05. Cells were equilibrated in motility buffer for \SI{30}{\minute}, and their swimming behavior was verified in a motility chamber prior to experiments. All microfluidic measurements were conducted within \SI{2}{\hour} at room temperature to maintain stable motility.

    \subsection*{Antibiotic treatments}
    
    To induce filamentation in susceptible cells, wild-type cultures were exposed to antibiotics after \SI{3}{\hour} of daytime growth. Cultures were then supplemented with either \SI{120}{\micro\gram\per\milli\liter} cephalexin or \SI{2}{\micro\gram\per\milli\liter} ampicillin and incubated for an additional \SI{2}{\hour} before experiments. This resulted in a total daytime culture duration of \SI{5}{\hour}, matching the incubation time used for untreated control cultures. An ampicillin-resistant \ecoli\ EPB47 strain harboring plasmid-encoded $\beta$-lactamase was used to study resistance effects.
    
    \subsection*{Microfluidic devices}
    
    The microfluidic channels were constructed with a depth of $H = \SI{10}{\micro\meter}$, a width of $W=\SI{50}{\micro\meter}$, and a length of several millimeters. To ensure a slow, steady, and stable flow rate, a resistor channel of approximately $\SI{10}{\centi\meter}$ in length was added. 
    The microfluidic channels were made of polydimethylsiloxane (PDMS) through the replication of a positive-relief silicon wafer master with SU-8 coated pattern fabricated by standard photolithography and soft-lithography procedures. The PDMS and the curing agent were mixed thoroughly at a 10:1 weight ratio. The mixture of PDMS and the curing agent was degassed inside a vacuum chamber for at least one hour to remove air bubbles before being poured onto the silicon wafer. The mixture was cured at $\SI{65}{\celsius}$ overnight to ensure full cross-linking. After cutting and peeling off the PDMS replica from the silicon wafer, $\SI{1}{\milli\meter}$ diameter holes were punched at the inlets and outlets. The PDMS replica and a glass coverslip were cleaned with compressed air and plasma-treated for 25 seconds (Harrick Plasma PDC-32G). We then irreversibly bound the PDMS replica and coverslip and left the sample on a $\SI{95}{\celsius}$ hot plate for $\sim \SI{1}{\minute}$ to strengthen the sealing. Flow was generated by applying a precise pressure gradient using a pressure-driven pump (ElveFlow OB1 MK4). 
    
    \subsection*{Imaging and single-cell tracking}
    
    We captured bacterial dynamics on a Nikon TI2-E microscope through a 20x objective (CFI60 Plan Apochromat Lambda Objective Lens, numerical aperture 0.75, working distance $\SI{1.0}{\milli\meter}$) under bright-field illumination for single-cell tracking and by fluorescence imaging for distribution analysis. The recordings were made with a scientific complementary metal–oxide–semiconductor (sCMOS) camera (Hamamatsu Orca-Fusion Gen-III) up to 100 FPS. 
    
    We flushed the microfluidic channels with bacteria suspended in motility buffer until an optimal cell concentration was reached that was dilute enough to prevent cell overlap and interference with tracking, yet sufficiently concentrated to obtain a robust dataset. Single-cell tracking was conducted to investigate breakout dynamics across different geometries under varying flow strengths.
    
    We consistently tracked cells swimming inside the channel without them moving out of the focal plane. Videos were post-processed in ImageJ, and cells were tracked automatically using a custom MACRO code with the help of the TrackMate plugin. Cell trajectories lasting longer than $\SI{1}{\second}$ were analyzed with a custom Python code to determine the breakout and infiltration probabilities and the upstream velocity distributions.

    \subsection*{Rheotaxis simulations}
    To isolate the effect of cell elongation on upstream migration, we simulated individual \textit{E. coli} cells swimming in a rectangular microchannel with width \(W=\SI{50}{\micro\meter}\) and height \(H=\SI{10}{\micro\meter}\). The simulations used a three-dimensional rheotaxis model with the analytical solution for pressure-driven Poiseuille flow in a rectangular channel. Each bacterium was modeled as a self-propelled cell body with position \(\mathbf{r}\), orientation \(\hat{\mathbf{p}}\), intrinsic swimming speed \(V_{\mathrm{swim}}\), aspect ratio \(\Gamma\), and flagellar length \(L_f\). Short wild-type-like cells were assigned aspect ratios uniformly distributed over \(\Gamma=2\text{--}4\), corresponding to cell lengths of \SI{2}{\micro\meter}--\SI{4}{\micro\meter} for a fixed cell width of \SI{1}{\micro\meter}, consistent with experimental measurements. Elongated cells were simulated by changing only the cell-body aspect ratio to \(\Gamma=8\text{--}10\), while keeping all other parameters fixed. The model includes intrinsic swimming, advection by the imposed flow, Jeffery-orbit rotation in shear flow, stochastic rotational noise, surface alignment, circular swimming near walls, weathervane reorientation, and corner interactions. Simulations were initialized with \(10^3\) bacteria randomly distributed in the channel and integrated with a time step of \(\delta t=\SI{0.01}{\second}\) for 5000 time steps. Full equations and parameter definitions are provided in Supplementary Notes S8 and S9.

\section*{Supplementary information}

Supplementary information (SI) is available.

\section*{Author contributions}

Ran Tao: conceptualization, methodology, investigation, device fabrication, data acquisition, formal analysis, software, simulations, visualization, writing -- original draft, and writing -- review \& editing. Nathaniel C. Esteves: resources, methodology, and writing -- review \& editing. Jay Zhu: resources, supervision, and writing -- review \& editing. Arnold J. T. M. Mathijssen: conceptualization, supervision, project administration, funding acquisition, and writing -- review \& editing.

\section*{Conflicts of interest}
The authors declare no conflict of interest.

\section*{Data availability}
All data supporting the findings of this study are available in the article. Additional information required to reanalyze the data is available from the corresponding author upon request.

\section*{Acknowledgements}

We thank all members of the Mathijssen lab and the Zhu lab for their support and insightful discussions. We further thank Liuni Chen and Ling Li for their support with SEM imaging. R.T. acknowledges support from the Dissertation Completion Fellowship at the University of Pennsylvania. A.J.T.M.M. acknowledges funding from the Charles E. Kaufman Foundation (Early Investigator Research Award KA2022-129523; and New Initiative Research Award KA2024-144001), the National Science Foundation (Career Award CBET-2542731; and UPenn MRSEC DMR-2309043), the University of Pennsylvania (URF, CURF, VIPER, Vagelos MLS, and FERBS programs), and the Research Corporation for Science Advancement (Cottrell Scholar Award CS-CSA-2026-125).

\balance

\bibliography{References_Antibiotics,references}
\bibliographystyle{rsc} 

\input{Figures}

\clearpage

\end{document}

%% file: Figures.tex
\begin{figure*}
    \includegraphics[width=\linewidth]{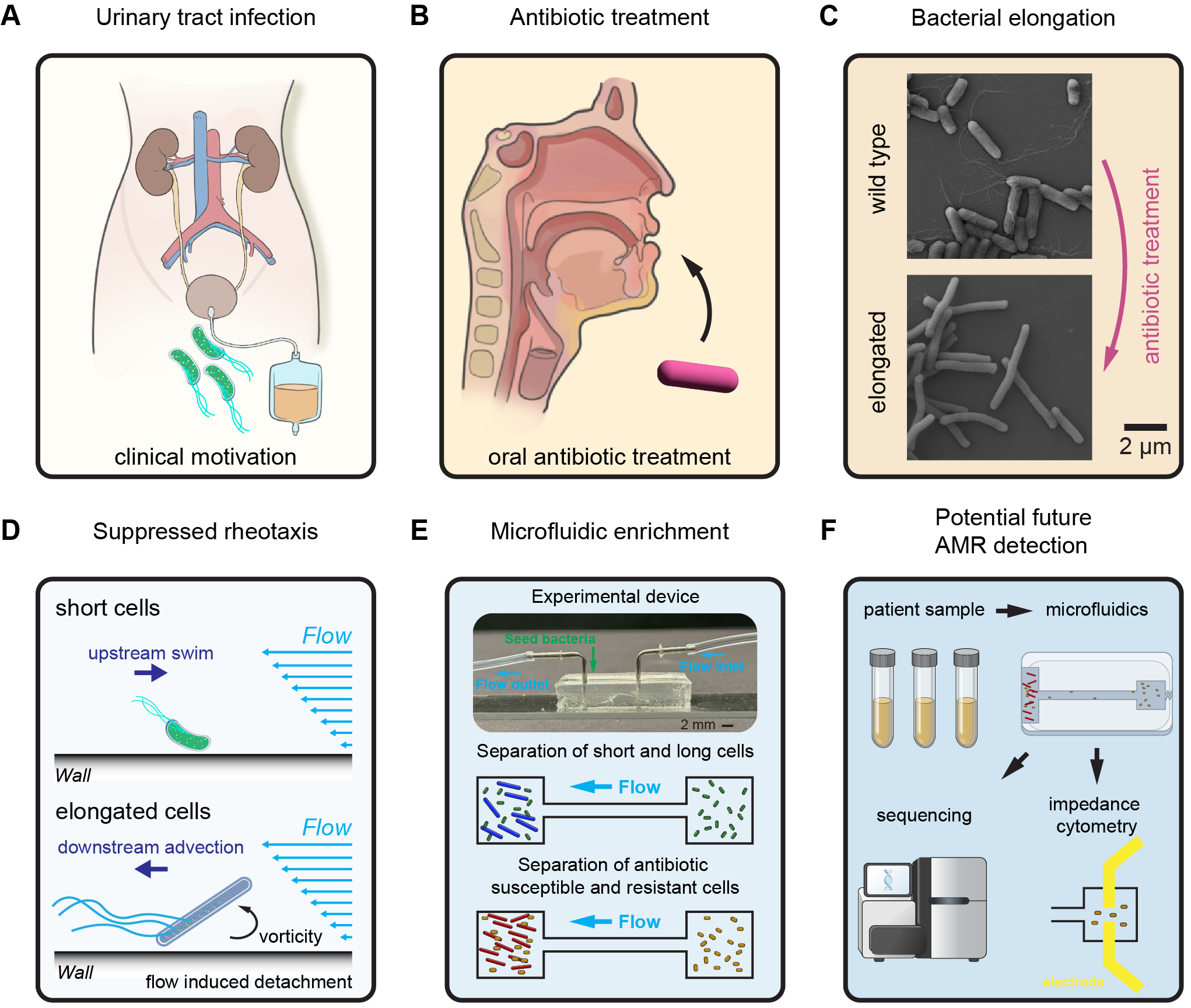}
    \caption{
    \label{fig1}
    \textbf{Overview of the experimental framework, physical mechanism, and potential diagnostic extension.}
    (A) Clinical motivation: motile bacteria can migrate upstream in urinary tract flows, motivating the investigation of bacterial transport under antibiotic exposure.
    (B) Oral antibiotic treatment introduces antimicrobial agents into the urinary tract.
    (C) $\beta$-lactam exposure inhibits cell division and induces filamentous elongation in antibiotic-susceptible \textit{E.~coli}.
    (D) Short cells maintain stable near-surface rheotaxis and swim upstream against the imposed flow, whereas elongated cells are more susceptible to flow-induced rotation, surface detachment, and downstream advection.
    (E) This length-dependent transport response enables microfluidic separation of short and elongated cells and proof-of-concept enrichment of ampicillin-resistant cells from mixed resistant and susceptible populations.
    (F) A potential future diagnostic workflow could combine microfluidic enrichment with impedance cytometry for microscopy-free on-chip detection of enriched cells, followed by downstream sequencing or other molecular analysis.
    Panels A, B, and F illustrate the clinical motivation and potential future workflow, whereas Panels C--E summarize the experimentally investigated mechanism and enrichment principle.
    }
\end{figure*}

\begin{figure*}
    \includegraphics[width=\linewidth]{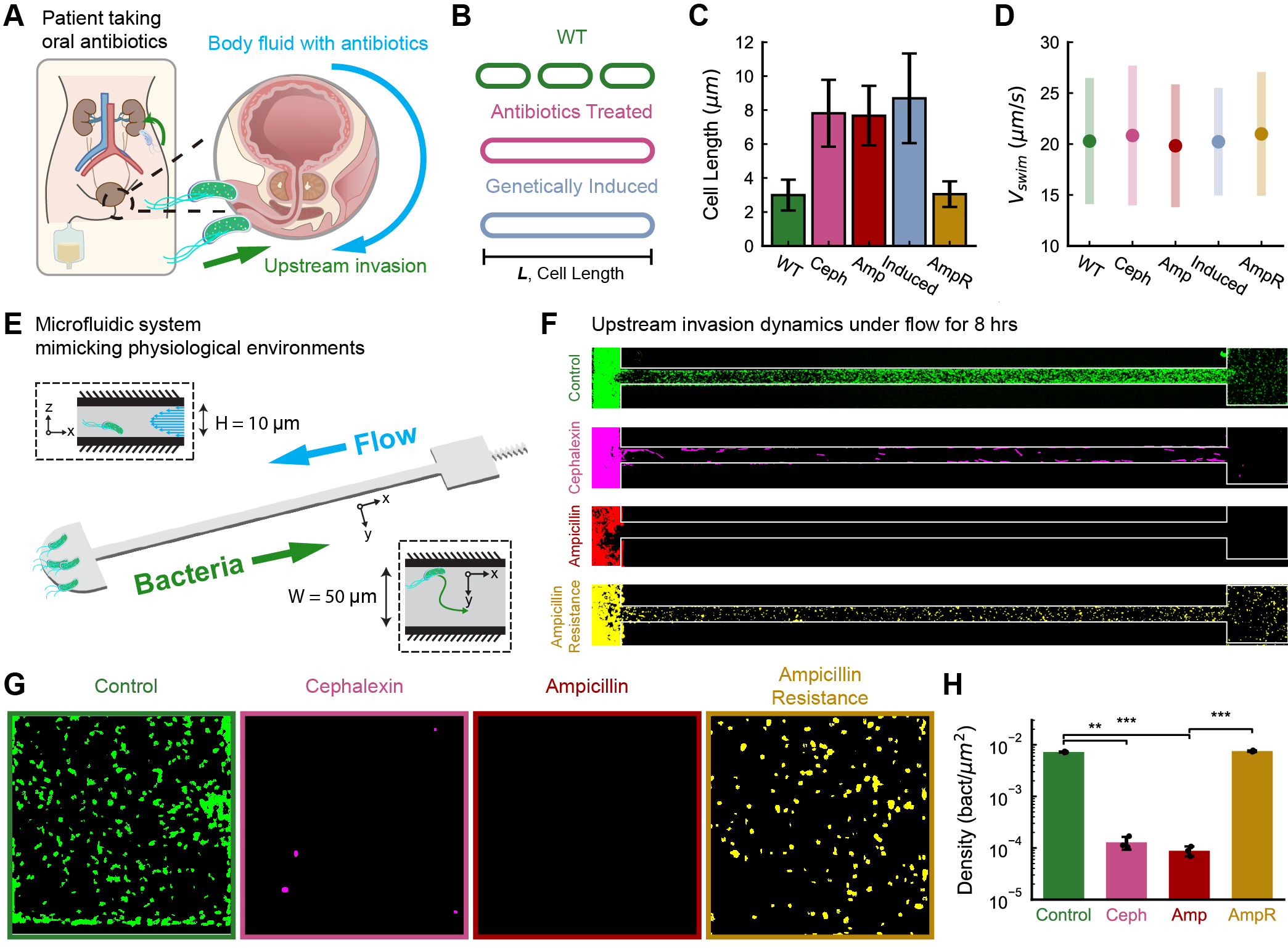}
    \caption{
    \label{fig2}
    \textbf{Antibiotic-induced elongation preserves motility but inhibits upstream invasion.}
    (A) Schematic of bacterial rheotaxis and antibiotic exposure in the urinary tract.
    (B) Representative morphologies: wild-type (WT) cells remain short rods, whereas cephalexin- and ampicillin-treated susceptible cells elongate. A genetically modified strain with inducible elongation is used to isolate the effect of cell length. An ampicillin-resistant strain retains short morphology under antibiotic treatment.
    (C) Quantification of cell length across conditions. Bars show the mean, and error bars indicate the standard deviation across $N=4$ independent cultures prepared on different days.
    (D) Swimming speeds, $V_{\text{swim}}$, across conditions. Points show the mean trajectory-level value, and shaded regions indicate the standard deviation across $N=4$ independent cultures prepared on different days.
    (E) Microfluidic platform used to quantify upstream invasion under imposed flow in channels of height $H=\SI{10}{\micro\metre}$ and width $W=\SI{50}{\micro\metre}$. Bacteria are introduced into the downstream reservoir, the narrow channel provides a simplified confined-flow environment motivated by bacterial transport in the urinary tract, and the upstream reservoir serves as the collection region for quantifying upstream migration.
    (F) Representative fluorescence images after \SI{8}{\hour} of flow for untreated WT, cephalexin-treated susceptible, ampicillin-treated susceptible, and ampicillin-resistant cells. White lines indicate the channel boundaries.
    (G) Representative images of the upstream reservoir after \SI{8}{\hour}.
    (H) Bacterial density in the upstream reservoir. Bars and error bars show the mean $\pm$ standard deviation across $N=4$ independent experiments; points show individual experiments. Asterisks indicate unadjusted $p$ values: *$p<0.05$, **$p<0.01$, and ***$p<0.001$.
    }
\end{figure*}

\clearpage
\begin{figure*}
    \includegraphics[width=\linewidth]{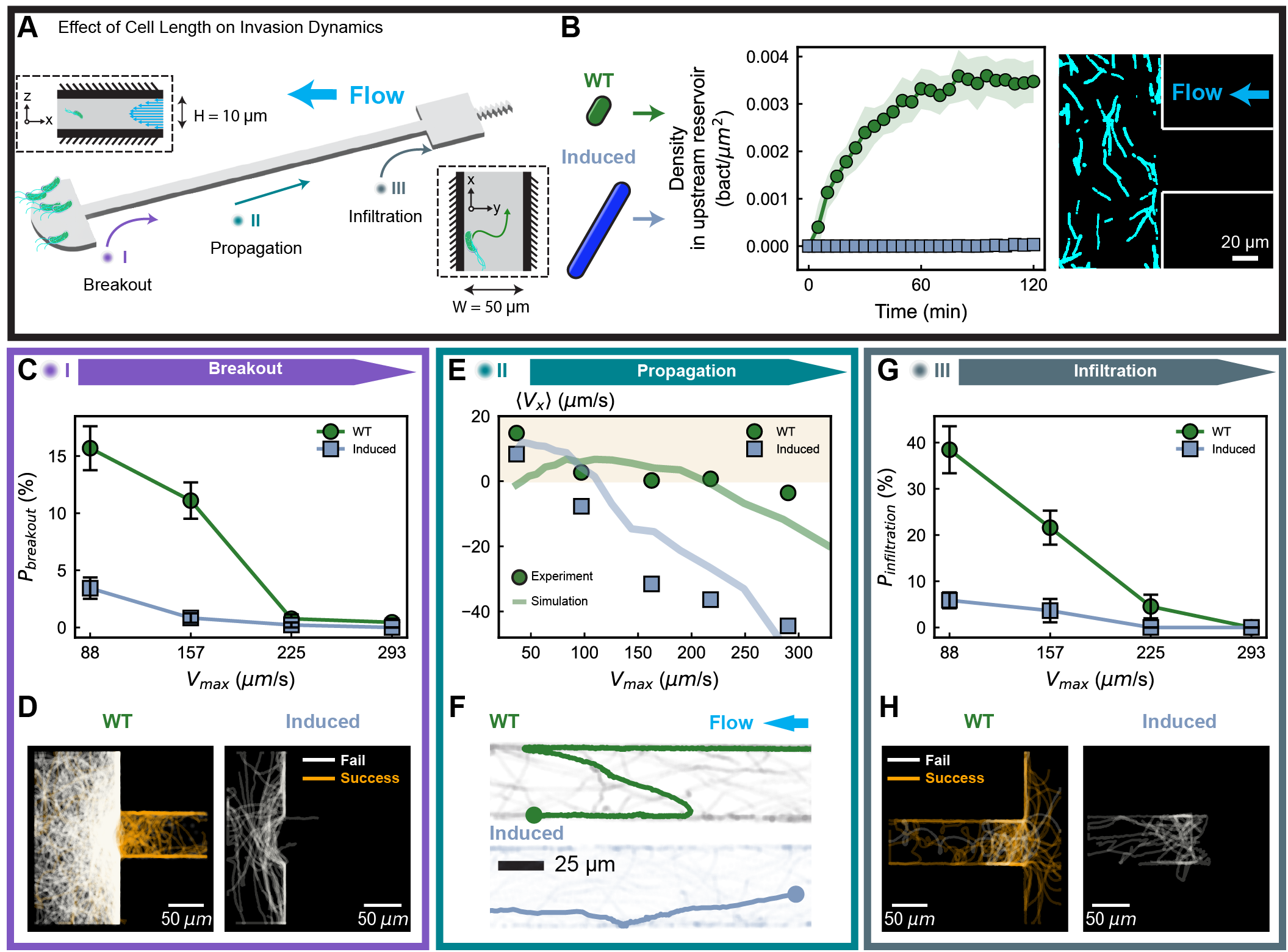}
    \caption{
    \label{fig3}
    \textbf{Elongated cells fail at distinct stages of upstream invasion.}
    (A) Microfluidic invasion device and three sequential stages: (I) breakout, (II) propagation, and (III) infiltration.
    (B) Time evolution of upstream invasion by wild-type cells (green) and elongated cells (blue) under flow.
    (C) Breakout probability, $P_{\text{breakout}}$, as a function of maximum flow velocity, $V_{\max}$. For each data point, $N>500$ breakout attempts were analyzed. Error bars represent the binomial standard error.
    (D) Representative breakout events for WT (left) and elongated cells (right). White and orange trajectories indicate failed and successful attempts, respectively.
    (E) Mean propagation velocity along the flow direction $\langle V_x \rangle$ versus $V_{\max}$. Dots represent experimental measurements, and solid lines represent simulation results.
    (F) Representative trajectories showing persistent edge-following and upstream motion in WT cells (top) and rapid detachment with downstream drift in elongated cells (bottom).
    (G) Infiltration probability, $P_{\text{infiltration}}$, into the upstream reservoir as a function of $V_{\max}$. For each data point, $N>500$ infiltration attempts were analyzed. Error bars represent the binomial standard error.
    (H) Representative infiltration events for WT (left) and elongated cells (right). White and orange trajectories indicate failed and successful attempts, respectively.
    }
\end{figure*}

\begin{figure*}
    \includegraphics[width=\linewidth]{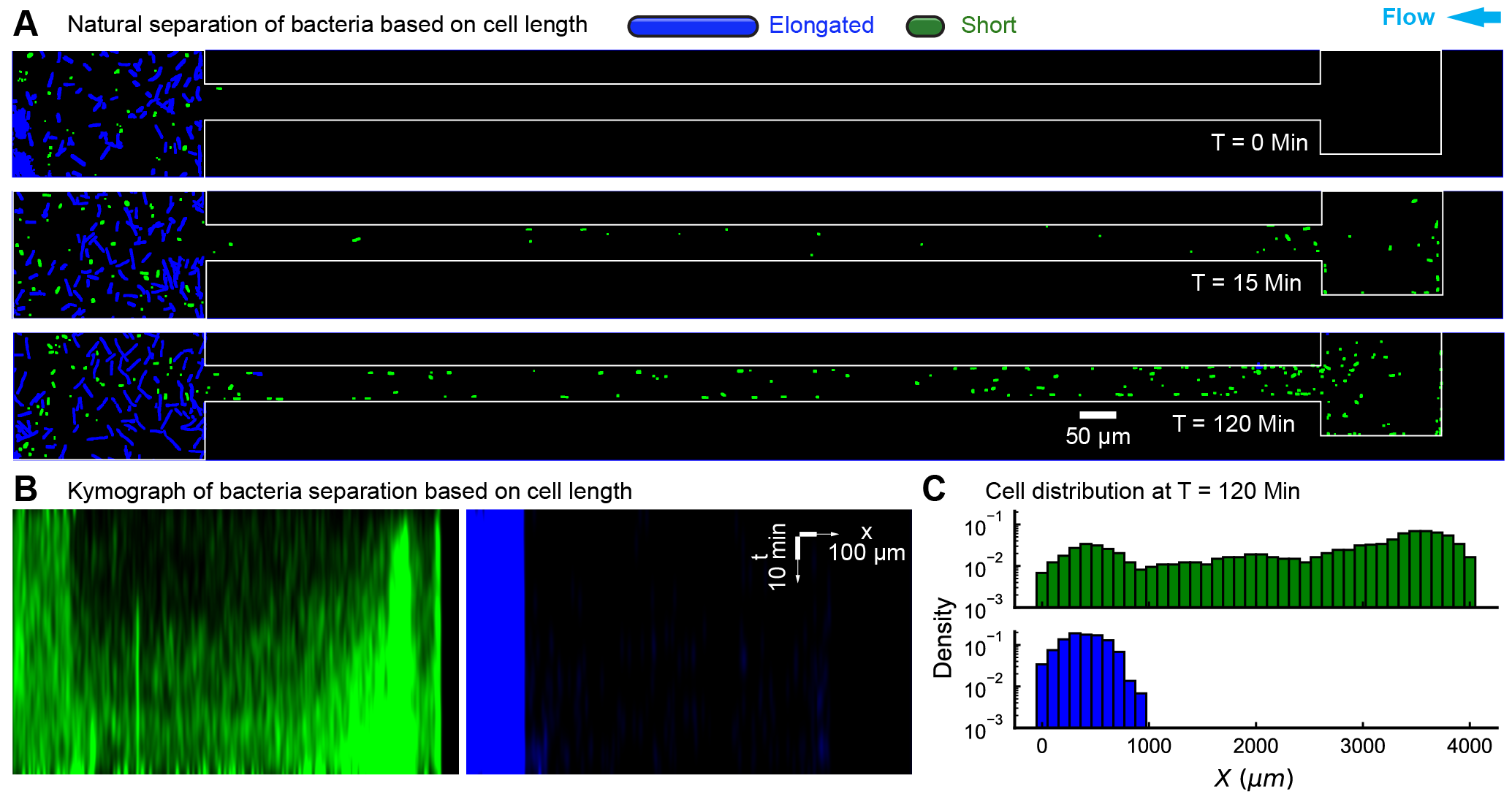}
    \caption{
    \label{fig4}
    \textbf{Flow-induced spatial separation of bacterial populations based on cell length.}
    (A) Time series of mixed short-cell populations (WT, green) and elongated-cell populations (blue) under imposed flow.
    (B) Space--time kymographs illustrating sustained upstream accumulation of short cells (left, green) and downstream retention of elongated cells (right, blue). Spatial position along the channel is shown on the horizontal axis and time progresses along the vertical axis.
    (C) Spatial density profiles after \SI{120}{\minute} show enrichment of short cells in the upstream region and depletion of elongated cells.
    }
\end{figure*}

\begin{figure*}
    \includegraphics[width=\linewidth]{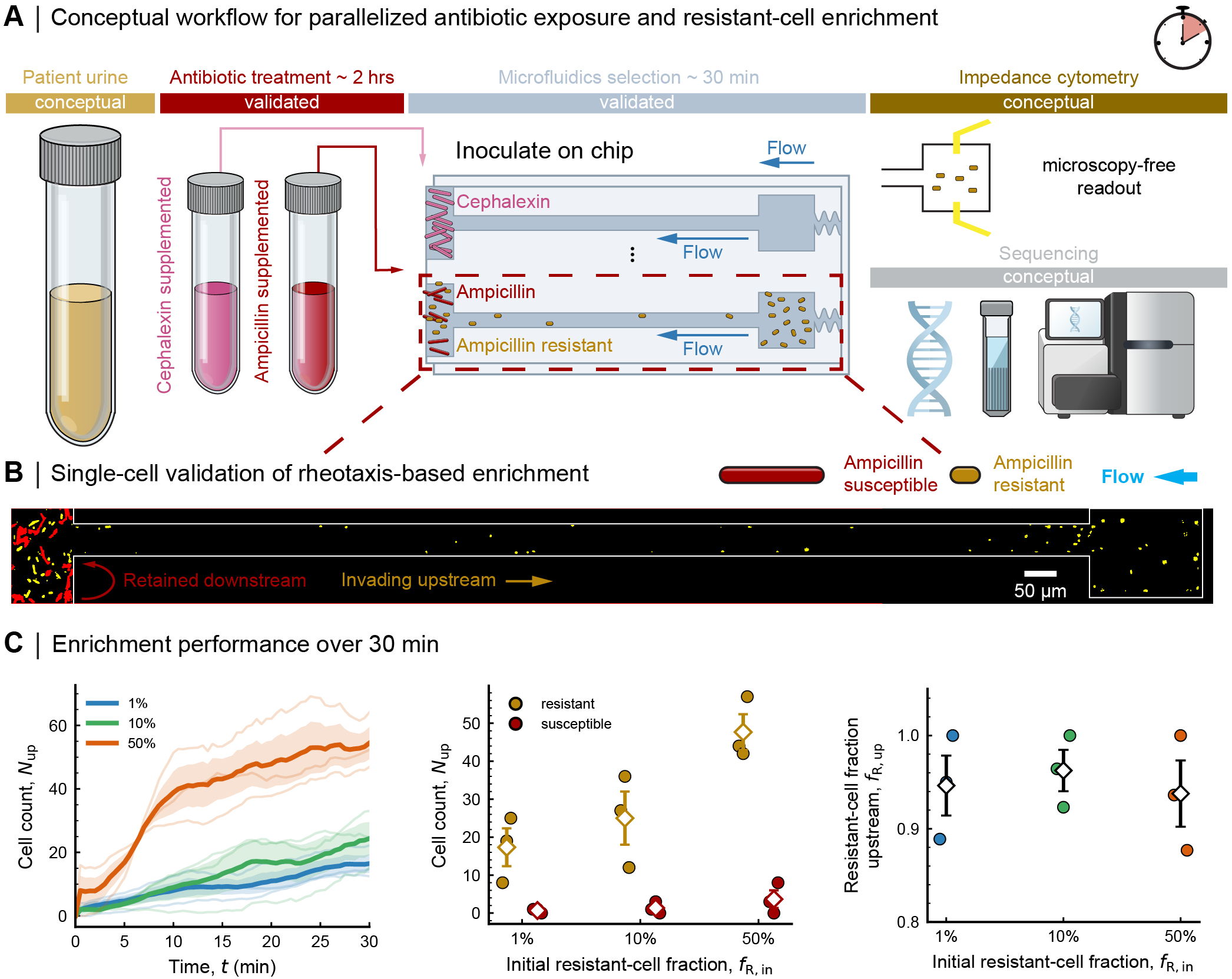}
    \caption{
    \label{fig5}
    \textbf{Rheotaxis-based microfluidic enrichment of ampicillin-resistant \ecoli.}
    (A) Conceptual workflow for antibiotic exposure, microfluidic enrichment, and downstream analysis. The dashed red box identifies the experimentally tested ampicillin condition. Patient-derived samples, parallel antibiotic testing, impedance cytometry, and sequencing are conceptual extensions.
    (B) Representative fluorescence image of ampicillin-resistant cells (yellow) accumulating upstream while susceptible elongated cells (red) remain predominantly downstream. The arrow indicates the flow direction. Scale bar, \SI{50}{\micro\meter}.
    (C) Enrichment performance for mixtures with different initial resistant-cell fractions. Left: total upstream cell count, $N_{\mathrm{up}}$, during the \SI{30}{\minute} enrichment period. Middle: resistant-cell and susceptible-cell counts upstream at \SI{30}{\minute}. Right: resistant-cell fraction upstream, $f_{\mathrm{R,up}}$, at \SI{30}{\minute}. Thin curves and circles show independent experiments; thick curves and open diamonds show the mean; shaded regions and error bars show mean $\pm$ SEM. For all conditions, $n=3$ independent experiments were conducted using separate samples on different days.
    }
\end{figure*}
